\documentclass[journal]{IEEEtran}
\usepackage{graphicx}
\usepackage{amsmath} 
\usepackage{amssymb} 

\begin{document}
\title{A Proton Simulator for Testing Implementations of Proton CT Reconstruction Algorithms on GPGPU Clusters\\}

\author{Micah~Witt$^{[1]}$,~\IEEEmembership{}
        Blake~Schultze$^{[1]}$,~\IEEEmembership{}
        Reinhard~Schulte$^{[2],[3]}$~\IEEEmembership{IEEE Member},
        Keith~Schubert$^{[1],[3]}$~\IEEEmembership{IEEE Sr. Member},
        and Ernesto~Gomez$^{[1]}$~\IEEEmembership{IEEE Member}

\thanks{Manuscript received November 16, 2012.  This work was supported by the National Institute of Health and the National Science Foundation under Grant No. 1R01EB013118-01 The content is solely the responsibility of the authors and does not necessarily represent the official views of the National Institute Of Biomedical Imaging And Bioengineering or
the National Institutes of Health.}
\thanks{[1]School of Computer Science and Engineering, California State University San Bernardino, San Bernardino CA 92407 USA, email: micah@r2labs.org.}%
\thanks{[2]Department of Radiation Medicine, Loma Linda University Medical Center, Loma Linda CA 92354 USA, email: rschulte@dominion.llumc.edu.}%
\thanks{[3]Department of Basic Sciences, Loma Linda University Medical Center, Loma Linda CA 92354 USA.}%
}

\maketitle
\pagestyle{empty}
\thispagestyle{empty}

\begin{abstract}
Proton computed tomography (pCT) is an image modality that will improve treatment planning for patients receiving proton radiation therapy compared with the current treatment techniques, which are based on X-ray CT.  Reconstruction of a pCT image requires solving a large and sparse system of linear equations, which should be accomplished within a few minutes in order to be clinically practical.  Analyzing the efficiency of potentially clinical reconstruction implementations requires multiple quality pCT data sets.  The purpose of this paper is to describe the simulator that was developed to generate realistic pCT data sets to be used in testing the efficiency of reconstruction algorithms, in particular string-averaging and block-iterative projection algorithms using sparse matrix formats on General Purpose Graphics Processing Units (GPGPU)s.
\end{abstract}


\section{Introduction}
\IEEEPARstart{P}{roton} computed tomography (pCT) is an imaging modality that tracks protons as they traverse objects to be imaged.  In proton radiation therapy, pCT imaging will provide improved treatment planning compared with the current method of X-ray CT imaging, because pCT will provide more accurate range prediction.  Accurate range prediction is important, for example, for sparing critical normal structures.

A pCT image is reconstructed by solving a very large system of linear equations of approximately 100 million equations with 10 million variables for a head-size object, in which the solution provides the relative stopping power (RSP) of every partition (voxel) of the object.  In order to solve such a system in a reasonable time frame so that pCT can be practical for clinical use, a parallel projection algorithm must be implemented across a multi-processor computing system.

Analyzing the convergence as well as the time and memory complexity of the reconstruction algorithms requires multiple high-quality data sets.  Generating real data in the laboratory by using the proton beam on a physical phantom is expensive and time-consuming.  Due to its highly technical nature, using the high energy physics simulation tool GEANT4 \cite{Agostinelli} to generate data sets is complex~\cite{Penfold}.  Therefore, a simulator that can quickly generate high-quality data sets will be essential in developing a clinically practical reconstruction methodology.  Furthermore, it will be advantageous to add several parameter options to assist in analyzing other reconstruction issues such as noise and path prediction.

This paper describes the simulator that has been developed to generate realistic data to be used in testing the efficiency of parallel reconstruction implementations on GPUs for pCT imaging.

\section{Methods}
A pCT simulator was developed to provide pCT data sets for solving the system of linear equations
\begin{eqnarray*} 
Ax=b
\end{eqnarray*}
where the $A$-matrix contains the proton path information, the $b$-vector contains the water equivalent path length (WEPL) of individual protons~\cite{Ford}, and the x-vector represents the unknown RSP values of each voxel. The simulator stores the actual-solution-vector for a given phantom.  The simulator creates a large user-determined number of realistic proton paths and corresponding elements of the matrix $A$, as well as the measurement vector $b$ of WEPL values with or without added noise.  The simulator can be operated in 2D or 3D mode providing the option of generating smaller 2D data sets along with realistically sized 3D data sets.  

\subsection{Phantoms}
The simulator developed in this work uses a general class of phantoms inspired by Herman's digital head phantom~\cite{Gabor} called a non-homogenious ellipse object (NEO) that allows for the addition of different RSP regions to represent different anatomical features.  

\begin{figure}
    \centering
    \includegraphics[width=2.4in]{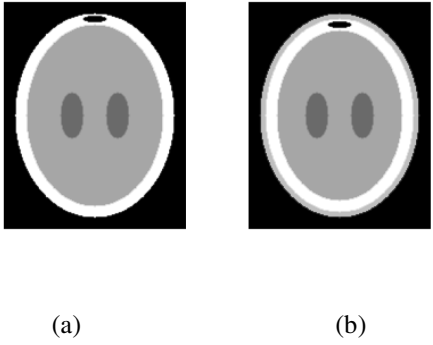}
    \begin{tabular}{cp{1in}c}
      (a) && (b) \\
    \end{tabular}
  \caption{ Non-homogenious Ellipse Objects (NEOs): (a) NEO~1, (b) NEO~2. }  
\label{Neo1and2}
\end{figure}

In this work two NEO phantom versions, NEO~1 and NEO~2 shown in Figure~\ref{Neo1and2} (a) and (b) respectively, were used to generate simulation examples.  Both phantoms are bounded by an outer ellipse with major and minor axis length of 180 mm and 140 mm, respectively, and have inner regions representing the cerebral ventricles filled with cerebro-spinal fluid (RSP of 0.9), brain tissue (RSP of 1.04), an air-filled frontal sinus (RSP of 0.0), and compact skull bone (RSP of 1.6).  NEO~2 also includes an outer skin layer. 

The simulator allows for different size voxels and reconstruction volume in 3D mode, or pixels and reconstruction area in 2D mode.  The 2D simulation examples used 1 mm x 1 mm pixels and a reconstruction area of 200 mm x 160 mm, consistent with the cross-sectional size of an adult human head, resulting in 32,000 pixels.  The simulator has three options for selecting the RSP for pixels along the elliptical boundaries containing two different RSP values.  The center-point method selects the RSP at the center of each pixel.  The corner-point averaging method uses the average of the RSP at the four corners of the pixel providing a smoother transition between different RSP regions than the center-point method.  The weighted-area average method provides the smoothest transition between two different RSP regions by setting the boundary pixel values to the sum of each region's RSP times the fraction of its area with respect to the pixel area.  The simulation examples used the corner-point averaging method.  

The simulator can also create 3D phantoms by stacking a user-defined number of slices with user-defined height of 2D phantoms, which results in an elliptical cylinder.  Stacking 200 1mm-slices of the phantom for example results in an elliptical cylinder within a reconstruction volume of 200x200x160 $= 6.4$ million voxels, where each voxel is 1 mm$^3$, creating data volumes consistent with the clinical setting of a human head.

\subsection{Proton Paths}
\begin{figure}
  \begin{center}
  \includegraphics[width=2.4in]{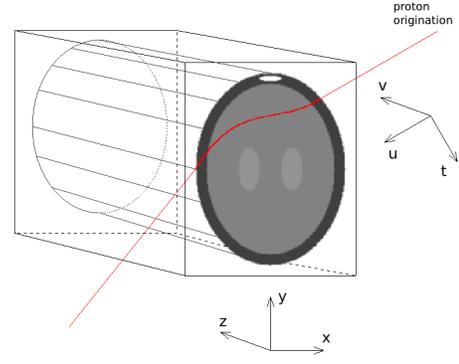}
  \caption{Proton simulator geometry. A randomly generated proton path passing through the digital phantom in the reconstruction space is shown.  The $t,v,u$ axes are the beam-specific coordinate system and the $x,y,z$ axes are the coordinates of the global reconstruction space.}  
  \label{ProtonPath}
  \end{center}
\end{figure}

The simulator projects parallel, linear proton paths from a virtual point source at infinite distance into the reconstruction space and then into the phantom.  The entry points into the reconstruction volume are uniformly distributed over a user-selected interval along the horizontal $t$-axis of the beam-specific coordinate sytem~(\ref{ProtonPath}) for the 2D simulation and along the horizontal axis and vetical $v$-axis for the 3D simulation.  In the simulation examples, were uniformly distributed over the interval [-125 mm, 125 mm] along the $t$-axis.

Once the entry point into the phantom is determined, the exit point is calculated by projecting a straight-line path with the addition of a lateral and angular displacement to model multiple Coulomb scattering (MCS).  Given the entry and exit points, the actual proton path connects the two points and then, outside the phantom, the path continues in a straight-line.  The simulator allows for either straight-line or cubic-spline paths.  The simulation examples used straight-line paths.

Proton paths are directed from a user-specified number of projection angles with a user-defined angular spacing interval.  The simulation examples used 180 projection angles with 2-degree spacing intervals, covering a full circle with a series of parallel 2D beams.  For each projection, the proton paths are generated in the $ut$-plane of the beam-specific coordinate system.

The random variables describing lateral and angular displacement at the proton exit are approximately distributed according to a bivariate normal distribution described by Highland's formalism of MCS~\cite{Highland} with constants recommended by the Particle Data Group~\cite{Nakamura}.  For the 2D mode of this simulator, a bivariate normal distribution is used to generate random number pairs to represent the exiting lateral displacement and angular deviation due to MCS of the proton within the phantom.  For the 3D mode, the random exit parameters are calculated for the horiozontal $ut$-plane and the vertical $uv$-plane separately since the lateral and vertical MCS of a proton are independent statistical processes.

For the bivariate normal distribution, let the two random variables be $X$ and $Y$, then the probability density function is
\begin{eqnarray*}
f(x,y)&=&\frac{1}{2\pi\sigma_x\sigma_y\sqrt(1-\rho^2)} exp\left(-\frac{1}{2(1-\rho^2)} \right.\\ 
&&\qquad\left[\frac{(x-\mu_x)^2}{\sigma_x^2} + \frac{(y-\mu_y)^2}{\sigma_y^2} \right. \\
&&\qquad\left.\left. - \frac{2\rho(x-\mu_x)(y-\mu_y)}{\sigma_x\sigma_y}\right]\right),
\end{eqnarray*}
where 
\begin{eqnarray*}
\rho&=&corr(X,Y)=\frac{v_{XY}}{\sigma_{xy}},\\ 
v_{XY}&=&cov(X,Y),\\ 
\mu&=&\left[ \begin{matrix} \mu_x  \mu_y \end{matrix} \right]\\
\Sigma&=&\left[ \begin{matrix}
      \sigma_x^2 & \sigma_{xy} \\
      \sigma_{xy} & \sigma_y^2  \\
\end{matrix} \right]\\
\end{eqnarray*}

The simulator has its own bivariate normal random number generator (BNRNG).  The BNRNG begins with two uniform random variables generated using the $rand()$ function in the C standard library.  It then converts those two uniform random variables to two standard independent random normals using the Marsaglia Polar-Method (modification of Box-Muller Method).  The two independent normals are then converted into a pair of bivariate normals using the covariance matrix.

\subsection{Mathematical Formulation of the Reconstruction Problem}
The elements of a given row of the $A$-matrix are the chord lengths of a given path through the pixels/voxels it intersected.  In the simulator, the element $a_{i,j}$ is the chord length of the $i^{th}$ path through the $j^{th}$ pixel/voxel.  For simplicity, the simulation examples used a constant chord length of the pixel size (1 mm).  

The path of the proton is generated in the $tvu$-coordinate system and then mapped to the stationary $xyz$-coordinate system using a Givens rotational matrix.  The simulator then takes the $xyz$ location of the path and intersects the path with the voxels of the reconstruction space.  Each proton path will intersect very few of the voxels in the entire reconstruction space so the vast majority of the entries in each row will be zero.  To identify every voxel intersected by a proton path, the simulator uses the Digital Difference Analyzer method~\cite{Gabor}.  

The noiseless WEPL value of the $i^{th}$ proton is calculated forming the inner product of the $i^{th}$ row with the actual-solution-vector of the phantom.  The simulator also has a feature that enables WEPL noise.  To do this, the noiseless WEPL value is converted into an equivalent energy value of the exiting proton using the NIST PSTAR database~\cite{Nist}.  Next, a random energy value is drawn from a normal distribution with mean equal to the noiseless energy and a standard deviation calculated using Tschalar's theory~\cite{Tschalar}.  This noisy energy value is then converted into the noisy WEPL.  

\begin{figure}
  \begin{center}
  \includegraphics[width=3.2in]{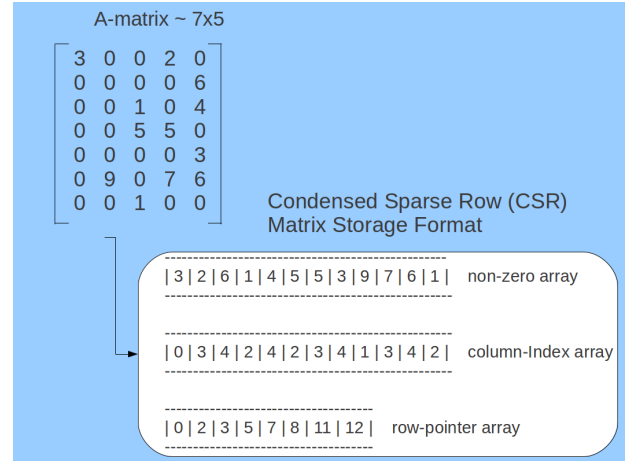}
  \caption{ An example of converting a dense matrix to condensed sparse row format.}  
  \label{CsrPic_Witt}
  \end{center}
\end{figure}

The number of voxels of the reconstruction volume defines the number of columns in the $A$-matrix.  A reconstruction volume of 200 mm x 160 mm x 200 mm bounds the typical adult human head.  A 3D simulation with a reconstruction volume of 6.4 million voxels and 64 million proton histories creates the system $Ax=b$ with $A \in \mathbb{R}^{64M \times 6.4M}$, $b \in \mathbb{R}^{64M}$.  The simulator writes the $A$-matrix to a file in a format that can be easily read into condensed sparse row (CSR) matrix format~\cite{BellGarland} taking approximately 100 GB of disk space (Figure~\ref{CsrPic_Witt}).  Storing the $A$-matrix in dense form would require over 1 peta-byte of disk space (1 peta $= 10^{15}$).  The data generated from a 2D simulation follows the same format.

\subsection{Reconstruction Algorithm and Hardware}

The string averaging projection (SAP) algorithm~\cite{Censor} was used in all reconstructions in this work.  SAP allows matrix row partitions called strings to be calculated in parallel.  The SAP algorithm is as follows:

Given the current iterate $x_k \in \mathbb{R}^n$, calculate $ \forall t \in \{1,2,...,M\}$, set $y^0=x^k$ and calculate, for $i=0,1,...,m(t)-1$,
\begin{eqnarray*}
y^{i+1}=y^i + \lambda_i\frac{<a^i,x_k> - b_i}{||a^i||_2^2}a^i,
\end{eqnarray*}
and let $y^t=y^{m(t)}$ for each $t=1,2,...,M$.  Then calculate the next iterate by
\begin{eqnarray*}
x^{k+1}=\sum_{t=1}^M w_ty^t,
\end{eqnarray*}
where for the weight vectors $w$, $\sum_{t=1}^M w_t=1$.  The reconstructions of the simulation examples were performed with $M=100$ string partitions.

\begin{figure}
  \begin{center}
  \includegraphics[width=4.4in]{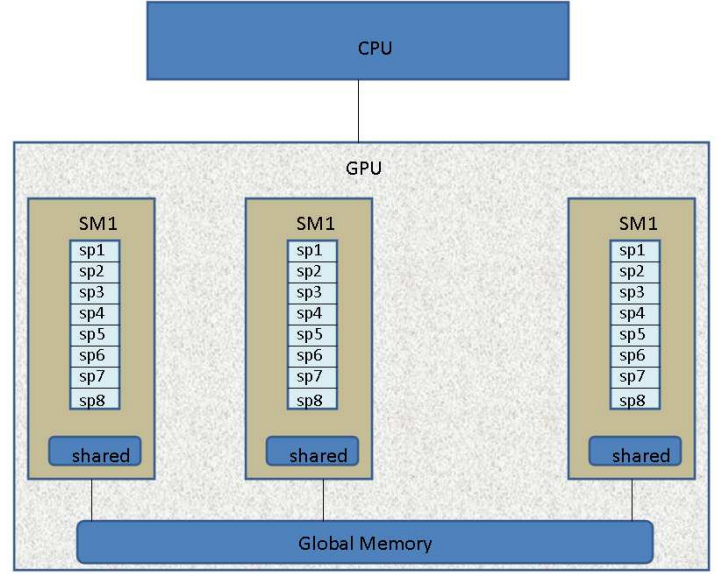}
  \caption{ GPU architecture: SM-symmetric multi-processor, SP-stream processor, mem-memory.  The GTX 280 has 30 SMs, with 8 SPs per SM for a total of 240 SPs, 1 GB of global memory, 16 KB of shared memory, 1.3 GHz clock speed. }  
  \label{GpuArch_Witt}
  \end{center}
\end{figure}

The SAP reconstruction algorithm was performed on a GTX 280 with 240 stream processors, 1GB global memory (Figure~\ref{GpuArch_Witt}).

\section{Results}
\begin{figure}
\begin{center}
\begin{tabular}{cc} 
  \includegraphics[width=1.1in,trim=0in 1in 6in 0in,clip=true]{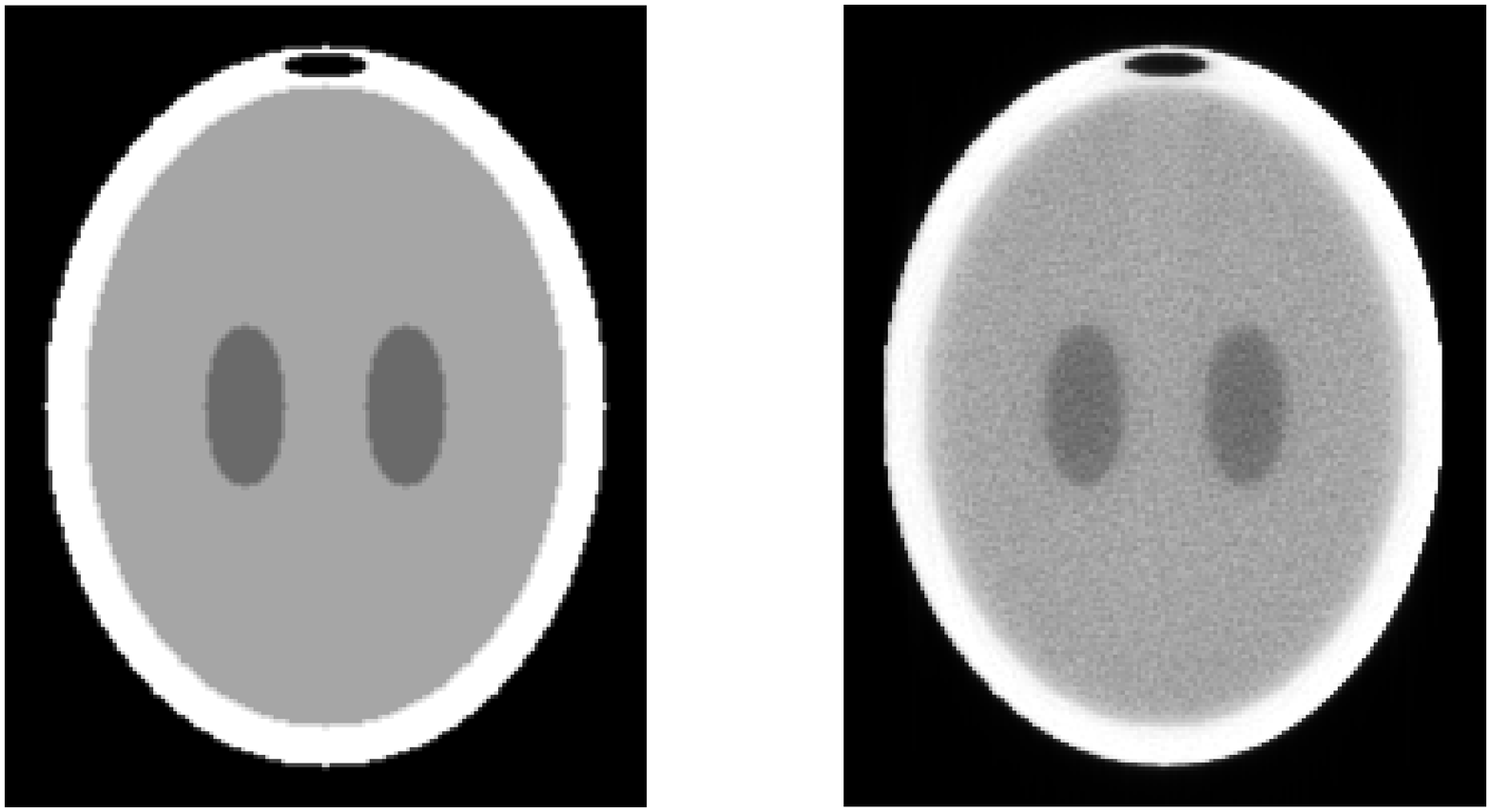}&  \includegraphics[width=1.1in,trim=6in 1in 0in 0in,clip=true]{pics/recon_Neo1_20it_10la.eps} \\
(a) & (b) \\
  \includegraphics[width=1.1in,trim=0in 1in 6in 0in,clip=true]{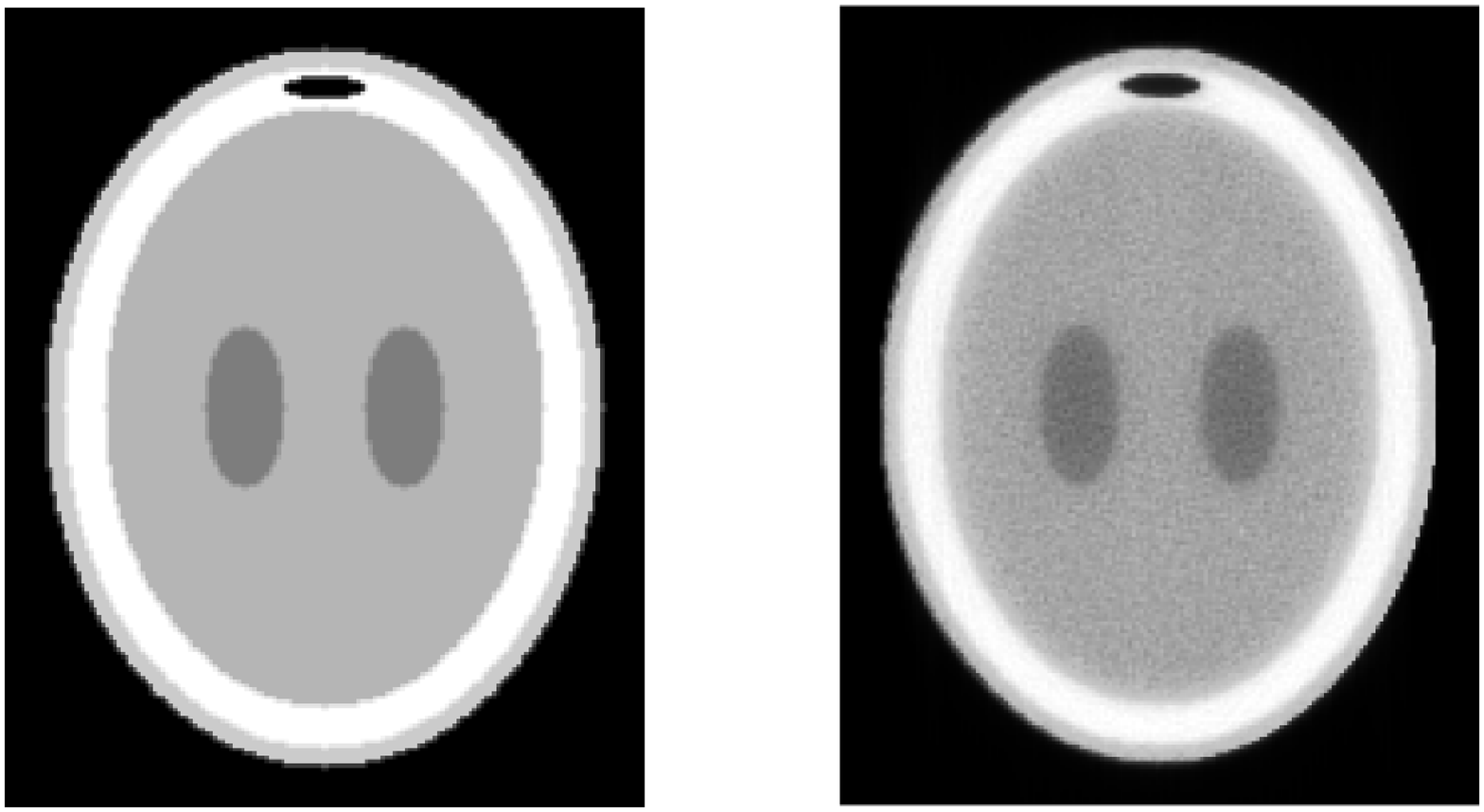}& \includegraphics[width=1.1in,trim=6in 1in 0in 0in,clip=true]{pics/recon_Neo2_20it_10la.eps}\\
(c) & (d) \\
\end{tabular}
  \caption{ Reconstructions ($n=200 \times 160$ pixels) of the NEO~1 and NEO~2 phantoms  ($\lambda=0.1$).  (a) NEO~1 phantom, (b) reconstruction of NEO~1 phantom, (c) NEO~2 phantom, (d) reconstruction of NEO~2 phantom. }  
\label{recon1and2}
\end{center}
\end{figure}

Figures 5 through 7 show examples of reconstructions obtained without added WEPL noise.  All reconstructions were performed with 20 iterations of the SAP algorithm.  Figure~\ref{recon1and2} illustrates the 2D reconstruction of two slightly different head phantoms, NEO~1 and NEO~2.  The reconstructions clearly show the different anatomical regions, characterized by different RSP values, for both phantom reconstructions. 

\begin{figure}
\begin{center}
\begin{tabular}{cc} 
  \includegraphics[width=1.1in,trim=0in 1in 6in 0in,clip=true]{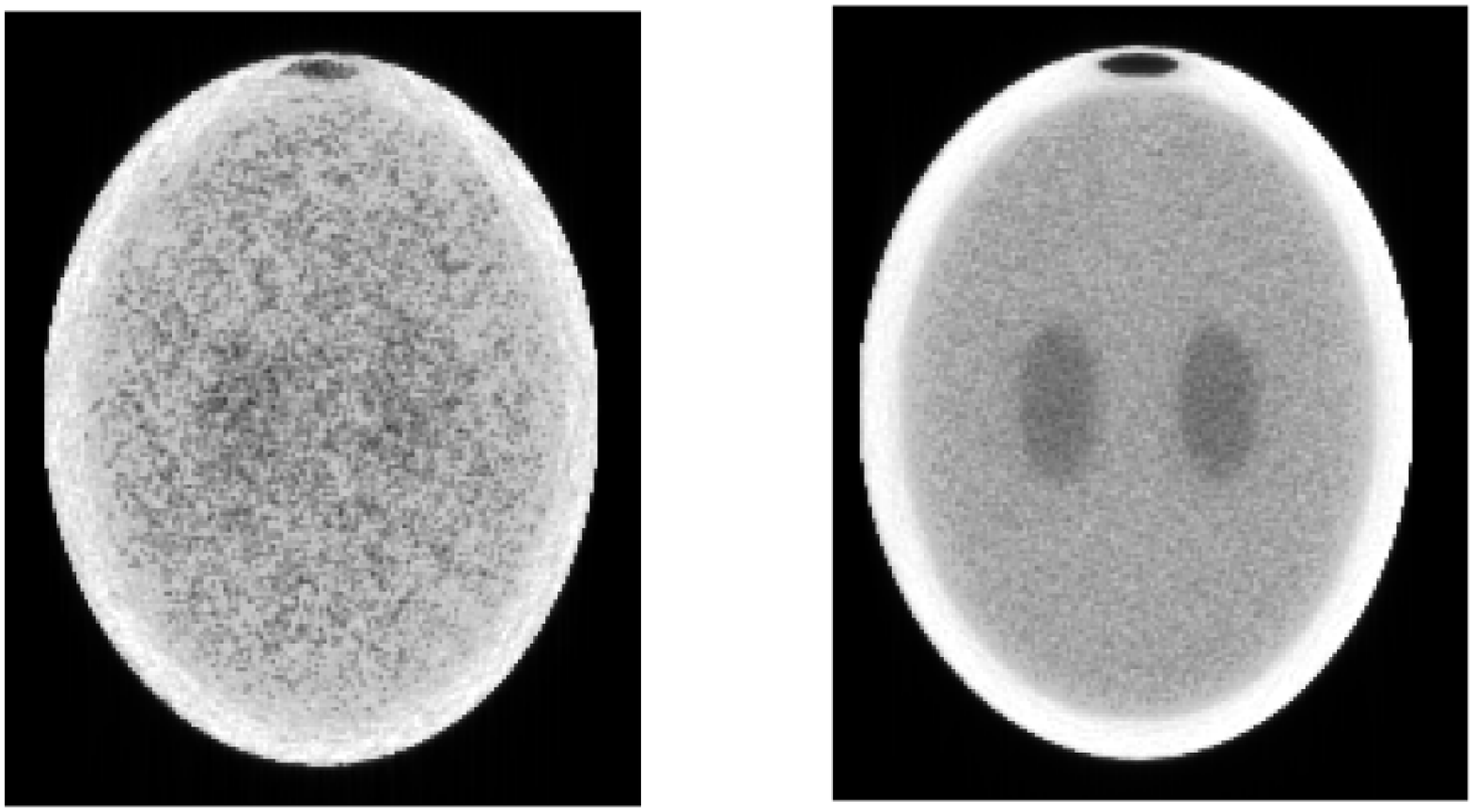}&  \includegraphics[width=1.1in,trim=6in 1in 0in 0in,clip=true]{pics/Fig6Top.eps} \\
(a) & (b) \\
  \includegraphics[width=1.1in,trim=0in 1in 6in 0in,clip=true]{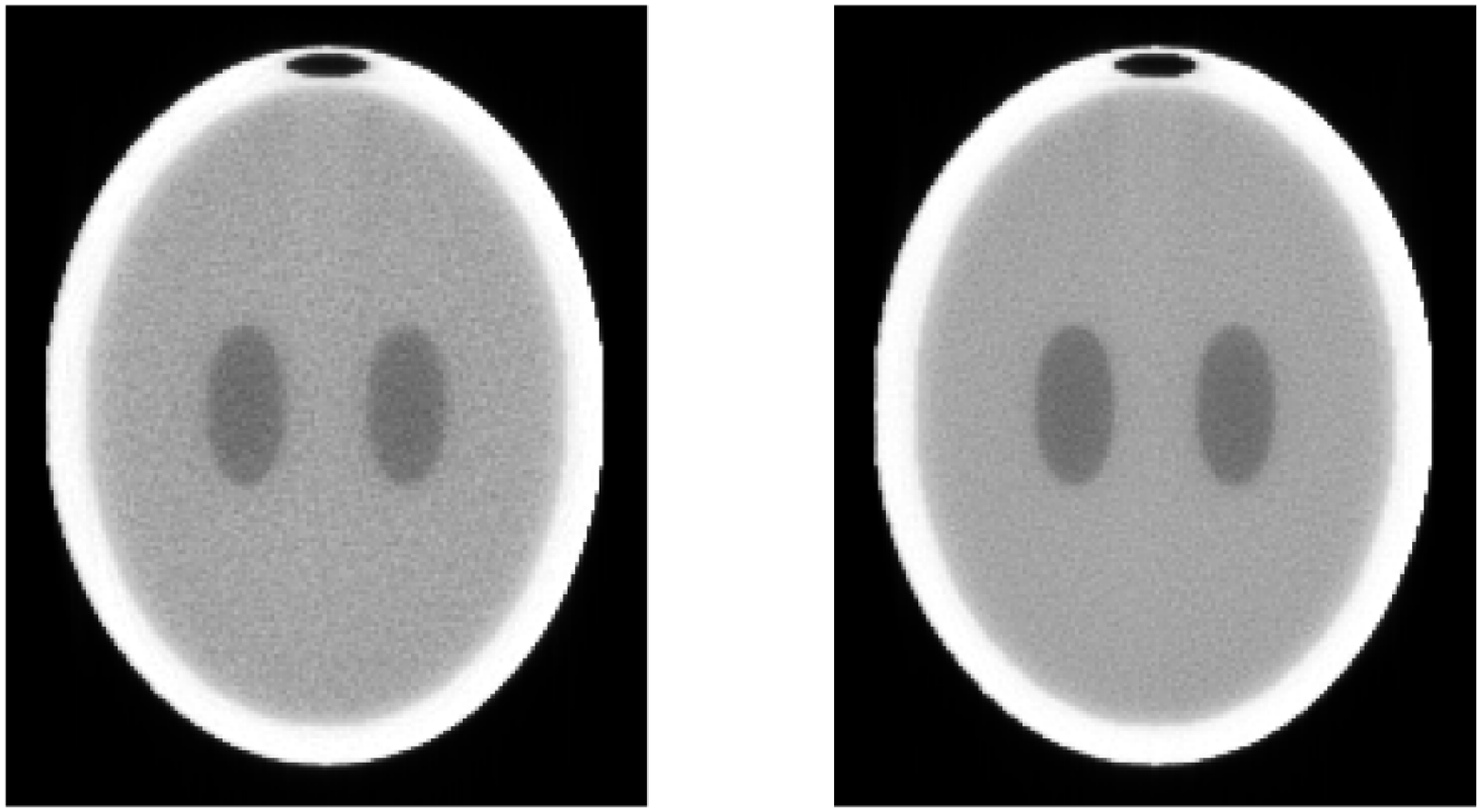}& \includegraphics[width=1.1in,trim=6in 1in 0in 0in,clip=true]{pics/Fig6Bot.eps}\\
(c) & (d) \\
\end{tabular}
  \caption{ Reconstruction of the NEO 1 phantom with different number of histories $m$, expressed as multiples of the number of pixels $n$ ( $\lambda=0.1$).  (a) $m=n$, (b) $m=5n$, (c) $m=10n$, (d) $m=20n$. }  
\label{recon3}
\end{center}
\end{figure}

Figure~\ref{recon3} analyzes the dependence of the image quality on the number of proton histories entering the reconstruction area, expressed as multiples of the number of pixels (32,000 pixels).  The resulting images show decreasing image noise with increasing number of histories.  The reason for the image noise, which is present despite the noiseless WEPL values, is the use of constant chord lengths in the reconstruction.  Obviously the noise introduced by inaccurate chord lengths can be compensated by a larger number of proton histories.

\begin{figure}
\begin{center}
\begin{tabular}{cc} 
  \includegraphics[width=1.1in,trim=0in 1in 6in 0in,clip=true]{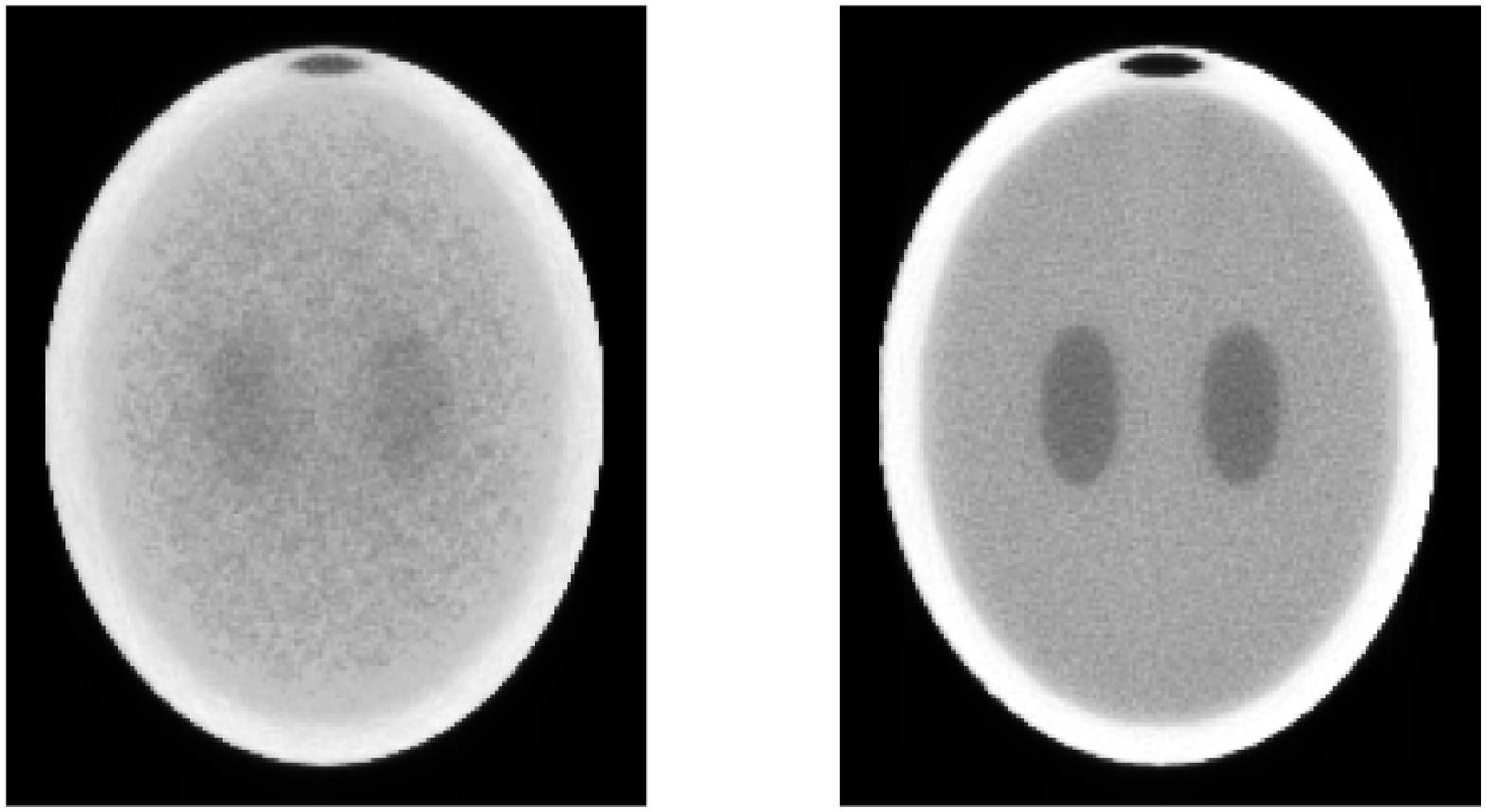}&  \includegraphics[width=1.1in,trim=6in 1in 0in 0in,clip=true]{pics/Fig7Top.eps} \\
(a) & (b) \\
  \includegraphics[width=1.1in,trim=0in 1in 6in 0in,clip=true]{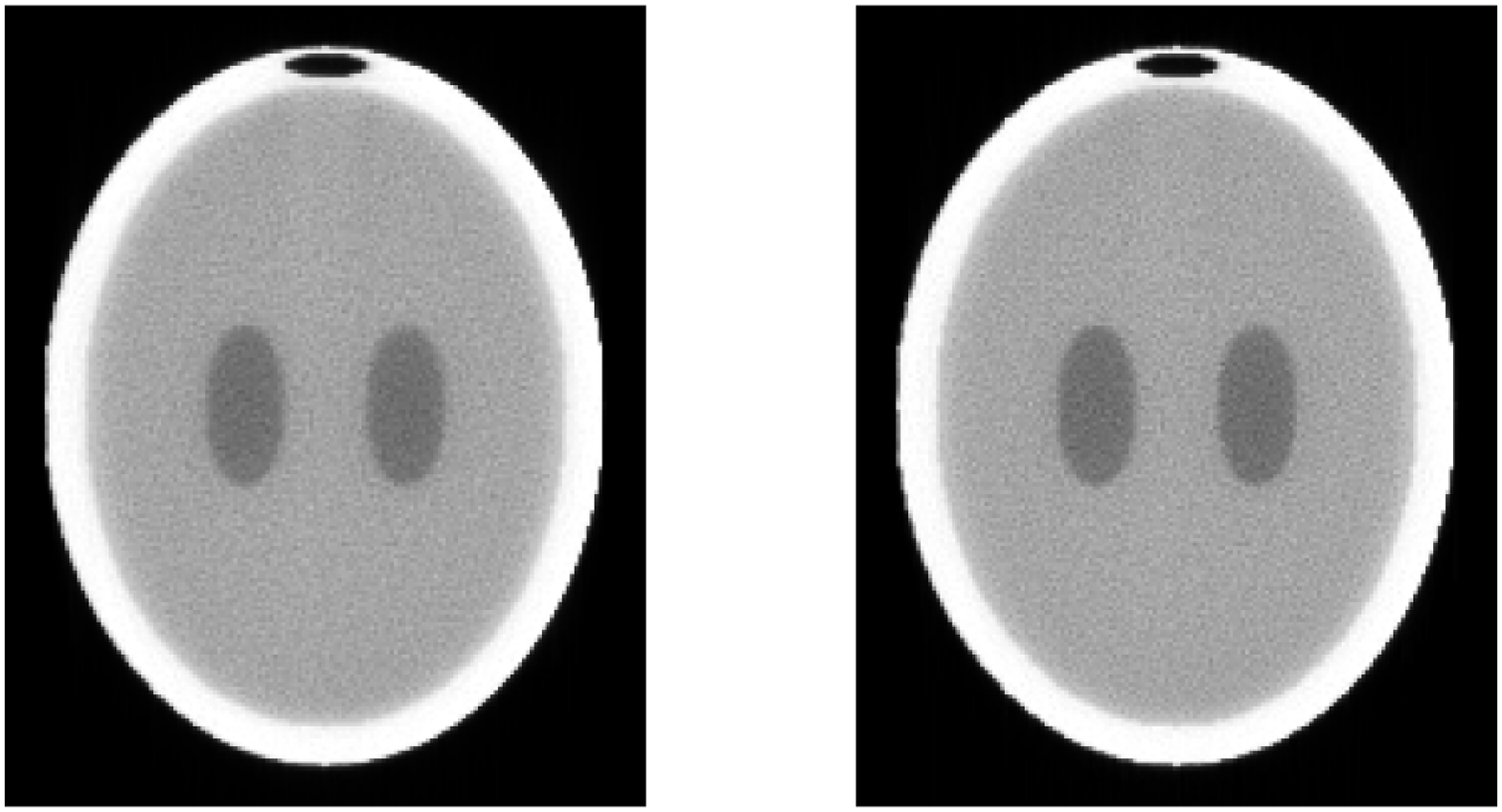}& \includegraphics[width=1.1in,trim=6in 1in 0in 0in,clip=true]{pics/Fig7Bot.eps}\\
(c) & (d) \\
\end{tabular}
  \caption{ Reconstructions of the NEO~1 phantom using different relaxation parameters $\lambda$. (a) $\lambda=0.01$, (b) $\lambda=0.1$, (c) $\lambda=0.2$, (d) $\lambda=0.5$. }  
\label{reconLa}
\end{center}
\end{figure}

Figure~\ref{reconLa} demonstrated the effects of different relaxation parameters on image quality with the same number of 20 iterations.  There is a decrease of image noise with an increase in $\lambda$, however, the effect of increasing $\lambda$ on noise reduction appears to saturate between $\lambda=0.2$ and $\lambda=0.5$.

\begin{figure}
  \begin{center}
  \includegraphics[width=3.8in]{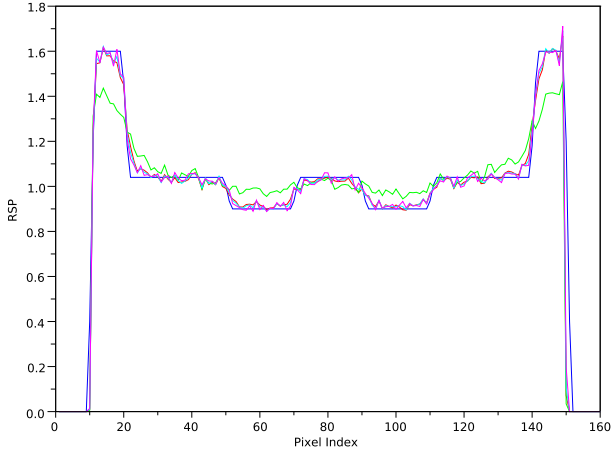}
  \caption{ Line profiles through the ventricles of the NEO~1 phantom for different relaxation parameters $\lambda$: NEO~1 (blue), $\lambda=0.01$ (green),  $\lambda=0.1$ (red), $\lambda=0.2$ (cyan), $\lambda=0.5$ (magenta).}  
  \label{LP_Lambdas}
  \end{center}
\end{figure}

\begin{figure}
  \begin{center}
  \includegraphics[width=3.8in]{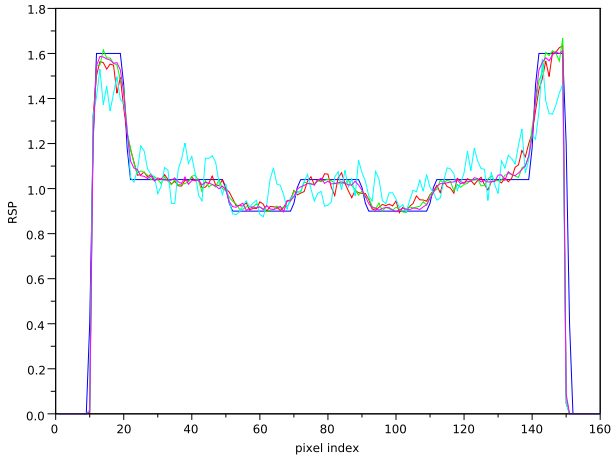}
  \caption{ Line profiles through the ventricles of the NEO~1 phantom using different numbers of histories: NEO~1 (blue), $m=n$ (cyan), $m=5n$ (red), $m=10n$ (green), $m=20n$ (magenta).}  
  \label{LP_Fig9}
  \end{center}
\end{figure}

Figures~\ref{LP_Lambdas} and \ref{LP_Fig9} demonstrate the effects of different numbers of histories and relaxation parameters on the quantitative accuracy of RSP values using line profiles through the two ventricles of the NEO~1 phantom.  Given a sufficient number of histories and choice of an adequate relaxation parameter, accurate reconstruction of the RSP values was obtained.

\section{Summary and Future Directions}

Fine-tuning and optimizing parameters of pCT reconstruction for the different varieties of pCT data that will be inherent in the actual application of this imaging modality requires a rapid turn-around of pCT reconstructions with realistic data sets.  The simulator presented in this work allows for pCT data sets to be created through a variety of options using a realistic head phantoms and transport parameters for protons.  It models a clinical setting in which virtual proton beams are directed from multiple angles simulating a 360-degree virtual proton gantry. Other features of the simulator, such as the ability to add noise or to use different path options with different accuracy, will be beneficial in systematically analyzing error sources of pCT reconstruction.
  
\begin{figure}
  \begin{center}
  \includegraphics[width=3.2in]{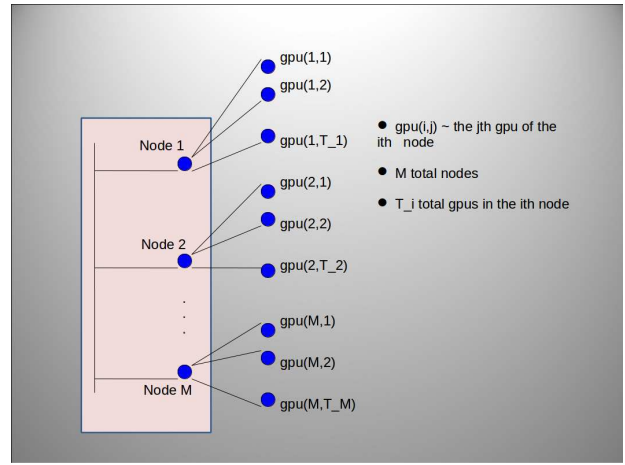}
  \caption{ Conceptual design of a GPU cluster with M nodes.  Each node has Ti GPUs, i=1,2,...,M.}
  \label{GpuClusterPic_Witt}
  \end{center}
\end{figure}

The data sets produced by the simulator are written to disk memory in a format that can be read into data structures that will later be used in the implementation of existing and novel parallel projection algorithms across GPGPU clusters.  Figure~\ref{GpuClusterPic_Witt} shows the conceptual design of a typical GPGPU cluster as a collection of nodes, with each node comprised of multiple GPUs.  

\begin{figure} 
  \begin{center}
  \includegraphics[width=3.2in]{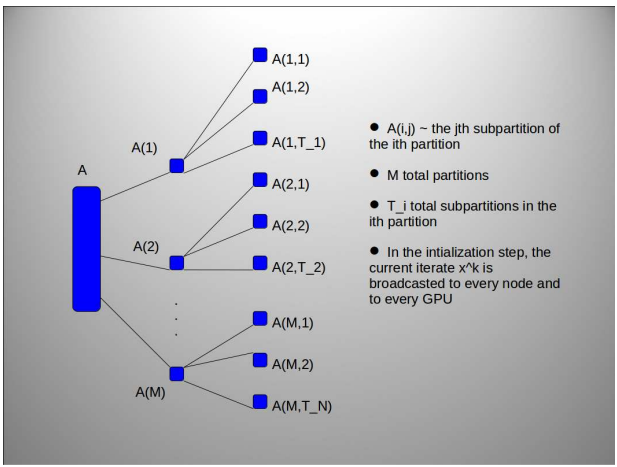}
  \caption{ General concept of matrix parititioning and assignment to GPUs in a GPGPU cluster. }
  \label{PartitioningPic_Witt}
  \end{center}
\end{figure}

One of the major goals of this work is to develop an $A$-matrix partitioning scheme that will assign these partitions to nodes within a cluster and to GPUs within a node matching the structure inherent in the spatial and temporal acquisition scheme of the pCT system and the physical nature of the data to the internal architecture of the GPGPU cluster efficiently, i.e., rendering as many GPUs active as possible and minimizing the need of data transfer between different nodes.  Figure~\ref{PartitioningPic_Witt} illustrates a general assignment scheme of matrix partitions to GPUs in a GPGPU cluster. The expected outcome of this research is a GPU-based reconstruction scheme that optimizes pCT reconstruction with respect to image quality, reconstruction time, and hardware expenses.

The pCT simulator, including source code and documentation, will be made readily available for researchers (physicists and engineers) as well as clinical physicists involved in proton therapy.  At this point, all simulation code has been written in C++, the graphics package utilizes the OpenGL libraries, and all data analysis code has been written in C++, MatLab, and SciLab.

\end{document}